\documentclass[conference]{IEEEtran}
\usepackage{float}
\usepackage{amsmath,amssymb,amsfonts}
\usepackage{graphicx}
\usepackage{cite}
\usepackage{algorithm}
\usepackage{algorithmic}
\usepackage{caption}
\usepackage{multirow, tabularx, booktabs, makecell}
\usepackage{amsthm}
\usepackage{hyperref}

\begin{document}

\title{Satellite Navigation: A Transmitting Intelligent Surface (TIS)-aided Indoor System}

\author{\IEEEauthorblockN{
Da Guan\IEEEauthorrefmark{1},
Xin Sun\IEEEauthorrefmark{1},
Tianwei Hou\IEEEauthorrefmark{1}\IEEEauthorrefmark{2},
Wenfei Gong\IEEEauthorrefmark{1},
Jun Wang\IEEEauthorrefmark{1},
Anna Li\IEEEauthorrefmark{3},
Arumugam Nallanathan\IEEEauthorrefmark{2}\IEEEauthorrefmark{4}
}\\
\IEEEauthorblockA{
\IEEEauthorrefmark{1} School of Electronic and Information Engineering, Beijing Jiaotong University, Beijing, CN\\
\IEEEauthorrefmark{2} School of Electronic Engineering and Computer Science, Queen Mary University of London, London, UK\\
\IEEEauthorrefmark{3} School of Computing and Communications, Lancaster University, Lancaster, U.K.\\
\IEEEauthorrefmark{4} Department of Electronic Engineering, Kyung Hee University, Yongin-si, Gyeonggi-do 17104, Korea
}}

\maketitle
\thispagestyle{empty}

\begin{abstract}
A transmitting intelligent surfaces (TISs) aided satellite indoor navigation system is investigated. By leveraging the unique features of TIS, we address the limitations of conventional global navigation satellite systems (GNSS) in providing reliable positioning services within indoor environments. To facilitate the extension of GNSS indoor signals, we establish an extended line-of-sight link using TIS which has the capability to change signal direction. A three-stage TIS-aided satellite indoor positioning algorithm (TSIPA), which utilizes the positions of TIS arrays and the angle of arrival, is proposed to locate indoor users. To evaluate the distribution of TIS arrays, we propose TIS position dilution of precision (TPDoP) to evaluate centroid deviation and utilize the root mean square error (RMSE) to represent compactness.
\end{abstract}

\section{Introduction}

The global navigation satellite system (GNSS) includes a global network of satellite navigation constellations, civil aviation augmentations and user equipment, which together provide all-weather, high-precision positioning services \cite{INT_GNSS1}. With the rapid advancement of sixth-generation mobile communication technology (6G), GNSS is applied in various fields \cite{INT_GNSS3}. However, the requirements for GNSS are increasingly strict, particularly in the high-precision indoor positioning \cite{INT_GNSS_indoor}. 

In traditional GNSS, navigation services primarily rely on medium-earth-orbit (MEO) satellites. After passing through obstacles such as wooden structures and brick walls, the signal power decays further to about $-160$ dBm in indoor scenarios. It is impossible to demodulate satellite signals in environments with numerous obstructions because of the low signal power\cite{INT_indoor2}. In addition, traditional GNSS systems typically require line-of-sight (LoS) links \cite{INT_LoS}. However, users do not have a direct LoS link to the satellite in indoor positioning scenarios within buildings. Recent research on satellite indoor positioning is primarily divided into two categories: 1) relay device based; 2) network based. However, these methods do not establish extended LoS links between satellites and indoor users and increase the system complexity greatly.

In order to solve the above mentioned problems of lacking LoS links, transmitting intelligent surface (TIS), which is an improvement of reconfigurable intelligent surfaces (RIS), is a promising solution. TIS can reconfigure the communication environment by changing the phase of the incoming signals \cite{INT_TIS1}. However, RIS can only provide services to the reflection space. Fortunately, simultaneously transmitting and reflecting (STAR)-RIS addresses the challenges of limited flexibility and effectiveness by enabling both transmission and reflection of incident signals \cite{INT_STARS1}. STAR-RIS is well-suited for assisting indoor positioning due to its ability to transmit signals to indoor user. Although these innovative studies have accelerated the development of 6G, the application of TIS in satellite navigation has not been fully explored yet. 
In addition, TIS is efficient for the lacking LoS links problem where there is no direct connection between the satellite and the user. \cite{INT_ELoS} investigates a RIS-aided GNSS network to enhance positioning services in urban canyons, which emphasizes the importance of optimal RIS placement for improving accuracy. To improve precision in ultra-long propagation paths with high path loss, \cite{INT_INAC1} proposes a non-orthogonal multiple access (NOMA)-RIS-enhanced MEO satellite network, integrating RIS to optimize power allocation for both communication and navigation. A hybrid communication and navigation network is proposed for balancing performance across different signal-to-noise ratio regimes. \cite{INT_INAC2} presents a RIS-aided INAC network which leverages RIS to increase received signal power in ultra-dense networks, thereby enhancing the navigation accuracy.

Based on the previous background, we propose a novel TIS-aided satellite indoor navigation system. By deploying multiple TIS arrays simultaneously, A three-stage TIS-aided satellite indoor positioning algorithm (TSIPA) is proposed to achieve pseudo-range positioning in carrier phase method (CPM) and carrier clock error method (CEM). In the first stage, we calculate the locations of the TIS arrays using the proposed improved pseudo-range measurement equations. In the second stage, maximum likelihood estimation (MLE) is used to estimate the AoA of the signal from TIS array to user. In the third stage, the location of indoor user can be calculated based on the locations of TIS arrays obtained in the first stage and AoA obtained in the second stage. We propose TIS position dilution of precision (TPDoP) as a new metric to evaluate the spatial distribution of TIS arrays. Furthermore, the function of positioning is achieved using satellite signals initially. Then, the subsequent positioning of user can be carried out only based on the stored TIS locations by indoor users.

\section{System Model}

\subsection{Channel model and signal model}
Considering a novel TIS-aided satellite navigation system in which $I$ satellites communicate with a single indoor user. Each satellite is equipped with $M$ transmitting antennas (TAs), while the user is equipped with $W$ receiving antennas (RAs). Both the satellite and the user's antenna are uniform line arrays (ULA). $R>1$ TIS arrays are configured on the windows of buildings to provide satellite navigation services for the indoor user. Each TIS array is equipped with $K$ elements. The indoor user is assumed to be randomly deployed in the indoor environment. The TIS arrays are deployed at the windows of the building, which receive signals from $I$ middle earth orbit satellites simultaneously, leading to the aliasing effects on the user's equipment. To mitigate this issue while maintaining general applicability, we activate $R$ time slots, where the time slots vector is given by $\left\{ {{T}_{1}}, \ldots, {{T}_{R}} \right\}$. It is guaranteed that only one TIS array is in operation in each time slot, thus providing the foundation for separating the signal of each TIS array. The coordinates of satellite ${{s}_{i}}$ are represented by $\left( {{x}_{i}},{{y}_{i}},{{z}_{i}} \right)$, where $i \in \mathcal{I} \buildrel \Delta \over = \left\{ {1,2, \ldots ,I} \right\}$. The coordinates of the TIS array $r$ can be simply defined as $\left( {{x}_{r}},{{y}_{r}},{{z}_{r}} \right)$, where $r \in \mathcal{R} \buildrel \Delta \over = \{ 1,2, \ldots ,R\}$. The coordinates of user is defined as $\left( {{x}_{u}},{{y}_{u}},{{z}_{u}} \right)$. Notably, both the locations of the TIS arrays and the indoor user are initially unknown. By employing geometric principles and satellite positions, we aim to accurately determine the positions of both TIS arrays and the indoor user.

The large-scale fading  between satellite $i$ and TIS array $r$ can be expressed as ${l_{ir}} = {G_T}{\left( {\frac{\lambda }{{4\pi {d_{ir}}}}} \right)^2}$, where ${G_T}$ represents the gain of transmitting antennas. $\lambda $ represents the carrier wavelength of the signal. $d_{ir}$ is the distance between satellite $i$ and the TIS array $r$. Similarly, the large-scale fading between TIS array $r$ and the indoor user can be given by ${l_{ru}} = {G_R}{\left( {\frac{\lambda }{{4\pi {d_{ru}}}}} \right)^2}$, where $d_{ru}$ represents the distance between TIS array $r$ and user. ${G_R}$ represents the gain of receiving antennas.

The small-scale fading vector between satellite $i$ and TIS array $r$ can be simply expressed as
\begin{equation}\label{small-scale-channel-satell-user}
{{\bf{h}}_{ir}} = {\left[ {\begin{array}{*{20}{c}}
{{h_{ir,11}}}& \cdots &{{h_{ir,1M}}}\\
 \vdots & \ddots & \vdots \\
{{h_{ir,K1}}}& \cdots &{{h_{ir,KM}}}
\end{array}} \right]_{K \times M}},
\end{equation}
where $k \in \mathcal{K} \buildrel \Delta \over = \{ 1,2, \cdots ,K\}$, $m \in \mathcal{M} \buildrel \Delta \over = \{ 1,2, \cdots ,M\}$. ${{h}_{ir,km}}$ is the small-scale fading between the $m$-th TA of satellite $i$ and the $k$-th element of the TIS array $r$, which experiences shadowed Rician fading \cite{shadowed_rice}.

The small-scale fading vector ${{\bf{h}}_{ru}}$ between TIS array $r$ and user is expressed as
\begin{equation}\label{small_scale_satel_RIS_column_i}
{{\bf{h}}_{ru}} = \left[ {\begin{array}{*{20}{c}}
{{h_{ru,11}}}& \cdots &{{h_{ru,1K}}}\\
 \vdots & \ddots & \vdots \\
{{h_{ru,W1}}}& \cdots &{{h_{ru,WK}}}
\end{array}} \right]_{W \times K},
\end{equation}
where ${h_{ru,wk}}$ denotes the small-scale channel fading between the $k$-th element of TIS array $r$ and the $w$-th RA of user, which also follows the shadowed Rician fading.

Given that TIS can transmit the incident signals throughout the transmission space, we define the transmission matrix of TIS array $r$ as follows:
\begin{equation}\label{transmission coefficient}
{{\bf{\Psi }}_r} = {\rm{diag}}\left( {{\beta _{1,r}}{\phi _{1,r}},{\beta _{2,r}}{\phi _{2,r}}, \cdots ,{\beta _{K,r}}{\phi _{K,r}}} \right),
\end{equation}
where ${\phi _{k,r}} = \exp (j{\theta _{k,r}})$ represents the phase shift of the transmitted signals by the ${k}$-th element of the TIS array $r$. $j = \sqrt { - 1} $.   ${\theta _{k,r}} \in \left[ {0,2\pi } \right)$ and ${\beta _{k,r}} \in \left[ {0,1} \right]$ denote the phase shift and amplitude corresponding to the transmission coefficients of the ${k}$-th element of TIS array $r$, respectively. By flexibly configuring the amplitude coefficient and phase shift of each TIS element, the signal can be superimposed in phase to enhance the signal amplitude.
Then, the signal received by the user from TIS array $r$ is given by
\begin{equation}\label{signal_starris}
{y_r} = \sum\limits_{i = 1}^I {\left( {{\bf{w}}{{\bf{h}}_{ru}}{{\bf{\Psi }}_r}{{\bf{h}}_{ir}}{{\bf{p}}_i}} \right)} \sqrt {{L_{ir}}{L_{ru}}{P_{\rm{T}}}} {s_{ir}} + {n_0},
\end{equation}
where ${P_{\rm{T}}}$ and ${s_{ir}}$ denote the transmitting power and the signal from satellite $i$ reaching the user via TIS array $r$, respectively. ${ {{{\bf{p}}_i}}}$ denotes the precoding vector of TAs of the satellite $i$ , while ${ {{{\bf{w}}}} }$ denotes the detection vector of RAs of the user. ${n_0}$ is the additive white Gaussian noise (AWGN).

\subsection{pseudo-range positioning model}
In this subsection, we introduce two scenarios: the CEM and the CPM. The difference between these two methods is that CEM determines the pseudo-range by multiplying the clock error by the speed of light while CPM determines the pseudo-range by multiplying the carrier phase difference by the wavelength. 

\subsubsection{\textbf {The positioning model under CEM}}

In satellite navigation, each satellite is equipped with an atomic clock, which can accurately record the time ${t_{si}}$ when satellite ${s_i}$ transmits a signal. The user records the time ${t_{u}}$ when the signal is received. Then, the distance between satellite ${s_i}$ and user through TIS array $r$ can be calculated by using the speed of light. When the electromagnetic waves are emitted from the satellite to the ground, atmospheric time delays and the inherent clock error between the user's clock and the precision atomic clock on the satellite introduce errors that significantly impact positioning accuracy. In a deterministic navigation system, it is worth noting that the clock error is the same for all satellites. Pseudo-range through CEM between satellite ${s_i}$ and user can be obtained by clock error, which is expressed as:
\begin{equation}\label{tranmit time}
\rho _{iu}^{{\rm{CEM}}} = c\left( {{t_u} - {t_{{s_i}}}} \right).
\end{equation}
where $c$ denotes the speed of light in vacuum.

Due to the lack of strict synchronization between the satellite clock, the user's clock and the positioning system's standard time, we assume that the clock error between satellite ${s_i}$ and the standard clock is $\Delta {t_{{s_i}}}$, as well as the clock error between the user's clock and the standard clock, denoted as $\Delta{t_{u}}$. Ignoring the time delays of the ionosphere and troposphere and other systematic delays, the actual distance between satellite ${s_i}$ to user can be further transformed into:
\begin{equation}\label{distance satellite user}
{d_{ir}} + {d_{ru}} = \rho _{iu}^{{\rm{CEM}}} - c\left( {\Delta {t_u} - \Delta {t_{{s_i}}}} \right),
\end{equation}
where $\Delta {t_{u}}$ is unknown. $\Delta {t_{s_i}}$ can be obtained by the navigation messages. ${d_{ir}}$ and ${d_{ru}}$ denote the Euclidean distances between satellite $s_i$ and TIS array $r$, and between TIS array $r$ and the user, respectively. Then, the actual distance between satellite ${i}$ and user can be simplified to
\begin{equation}\label{e119}
{d_{ir}} + {d_{ru}} = \rho _{iu,c}^{{\rm{CEM}}} - c\Delta {t_u}.
\end{equation}
where $\rho _{iu,c}^{{\rm{CEM}}} = \rho _{iu}^{{\rm{CEM}}} + c\Delta {t_{s_i}}.$

The positioning equation contains four unknowns ${x_u},{y_u},{z_u},{\Delta{t_u}}$. In practical applications, it is crucial to recognize that navigation necessitates the involvement of a minimum of four satellites. However, in indoor environments, a direct connection between the satellite and the user is unavailable. Fortunately, the TIS arrays can offer alternative signal paths, and the IPR positioning equations for TIS array $r$ can be formulated as follows:
\begin{equation}\label{e20}
\left\{ {\begin{array}{*{20}{c}}
{\rho _{1u,c}^{{\rm{CEM}}} = {d_{1r}} + {d_{ru}} + c\Delta {t_u}},\\
 \vdots \\
{\rho _{Iu,c}^{{\rm{CEM}}} = {d_{Ir}} + {d_{ru}} + c\Delta {t_u}}.
\end{array}} \right.
\end{equation}

\subsubsection{\textbf {The positioning model under CPM}}
The main idea of CPM is to calculate the distance through the carrier phase and wavelength. The carrier phase pseudo range consists of two parts: integer part $d_{int}$ and decimal part $d_{dec}$. To calculate ${{d}_{ int }}$, the number of whole period $N$ should be evaluated, which is also called the integer ambiguity. Then, ${{d}_{ int }} = \lambda N$. Next, the fractional part of the carrier phase pseudo-distance can be obtained by multiplying the wavelength by the proportion of the carrier phase difference in a cycle, ${d_{dec}} = {{\lambda \Delta \varphi } \mathord{\left/ {\vphantom {{\lambda \Delta \varphi } {2\pi }}} \right. \kern-\nulldelimiterspace} {2\pi }}$.

Similar to \eqref{distance satellite user} - \eqref{e20}, the corrected carrier phase difference is given by:
\begin{equation}\label{corrected_phi}
\Delta {\varphi _{iu,c}} = \lambda _i^{ - 1}\left( {{d_{ir}} + {d_{ru}} + c\Delta {t_u}} \right) - {N_i},
\end{equation}
where ${\lambda _i}$, $\Delta{\varphi _{iu,c}}$ and $N_i$ denotes the wavelength of the signal emitted by the $i$-th satellite, the carrier phase change of the $i$-th satellite signal and the integer ambiguity of the $i$-th satellite signal, respectively. Then, the observation equations of carrier phase method for TIS array $r$ can be given by:
\begin{equation}\label{CPM}
\left\{ {\begin{array}{*{20}{c}}
{\rho _{1u,c}^{{\rm{CPM}}} = {d_{1r}} + {d_{ru}} + c\Delta {t_u},}\\
 \vdots \\
{\rho _{Iu,c}^{{\rm{CPM}}} = {d_{Ir}} + {d_{ru}} + c\Delta {t_u}}.
\end{array}} \right.
\end{equation}
where the pseudo-distance through CPM can be given by $\rho _{iu,c}^{{\rm{CPM}}} = {\lambda _i}\left( {\Delta {\varphi _{iu,c}} + {N_i}} \right)$.

The integer ambiguity $N_i$ can be obtained through the LAMBDA method described in \cite{LAMBDA}. Then, it can be readily found that \eqref{e20} and \eqref{CPM} can be solved by using the same mathematical algorithm. 

\section{Position Algorithm and AoA Estimation} \label{section_3}

TSIPA is divided into three stages. The first stage involves treating the distance from the TIS to the user as a systematic error and calculating the coordinates of the TIS arrays using the least squares method. The second stage is to estimate AoA of the signal from TIS arrays to user utilizing MLE. The third stage entails calculating the user location based on the TIS array locations and the signal's AoA, applying principles of solid geometry and mathematical optimization techniques.

\subsection{Positioning TIS arrays}

Since LSM can only solve linear equations, it is necessary to transform the problem into a linear form. We then define that $\rho _{iu,c}^{iter}$ represents the corrected pseudo-range between satellite ${s_i},i \in \left\{ {1,2,3,4} \right\}$ and initial solution, which is expressed as
\begin{equation}
\begin{aligned}
\rho _{iu,c}^{iter} 
&= d_{ir}^{iter} + d_{ru}^{iter} + c\Delta t_u^{iter},
\end{aligned}
\end{equation}
where $d_{ir}^{iter}$ and $d_{ru}^{iter}$ represent the Euclidean distances of $s_i$-TIS and TIS-user links in each iteration, respectively. 
The standard form of LSM is given by
\begin{equation}\label{equa0}
{\bf{b}} = {\bf{A}}{\bf{x}} + {\bf{n}},
\end{equation}
where ${{\bf{x}}_{4 \times 1}}$, ${{\bf{b}}_{I \times 1}}$, ${{{\bf{n}}_{I \times 1}}}$ and ${{\bf{A}}_{I \times 4}}$ denote the desired vector, observation vector, noise vector and system transformation relation matrix, respectively. The expression of each symbol can be expressed as:

\begin{equation}\label{b_in_eq0}
{\bf{b}}\left[ i \right] = \rho _{iu,c}^{{\rm{CEM/CPM}}} - \rho _{iu,c}^{{\rm{CEM/CPM}},iter},
\end{equation}
\begin{equation}\label{A_in_eq0}
{\bf{A}}\left[ {i,:} \right] = \left[ {\begin{array}{*{20}{c}}
{\frac{{{x_i} - {x_{r0}}}}{{{{\left\| {{{\bf{s}}_i} - {{\bf{x}}_{r0}}} \right\|}_2}}}}&{\frac{{{y_i} - {y_{r0}}}}{{{{\left\| {{{\bf{s}}_i} - {{\bf{x}}_{r0}}} \right\|}_2}}}}&{\frac{{{z_i} - {z_{r0}}}}{{{{\left\| {{{\bf{s}}_i} - {{\bf{x}}_{r0}}} \right\|}_2}}}}&C
\end{array}} \right],
\end{equation}
\begin{equation}\label{dx_in_eq0}
{\bf{x}} = {\left[ {\begin{array}{*{20}{c}}
{{x_r} - {x_{r0}}}&{{y_r} - {y_{r0}}}&{{z_r} - {z_{r0}}}&{\Delta {t_c} - \Delta {t_{c0}}}
\end{array}} \right]^{\rm{T}}},
\end{equation}
and
\begin{equation}\label{n_in_eq0}
{\bf{n}} = {\left[ {\begin{array}{*{20}{c}}
{{n_1}}& \cdots &{{n_I}}
\end{array}} \right]^{\rm{T}}},
\end{equation}
where ${{\bf{s}}_i}$ and ${{\bf{x}}_{r0}}$ denote the 3D coordinate vectors of the $i$-th satellite and the iteration starting point. ${\bf{A}} = \left[ {i,:} \right]$ denotes the all the elements of the i-th row of matrix ${\bf{A}}$. Constant $C$ is set to $1$. $\Delta {t_c}$ is the time it takes for the signal to travel from TIS array $r$ to the user, $\Delta {t_c} = d_{ru}/c$.
The cost function ${\bf{P}}({\bf{x}})$ is defined as
\begin{equation}\label{P(X)}
\begin{aligned}
{\bf{P}}({\bf{x}}) &= {({\bf{Ax}} - {\bf{b}})^{\rm{T}}}{\bf{W_e}}({\bf{Ax}} - {\bf{b}}),
\end{aligned}
\end{equation}
where $\bf{W_e}$ is the weight matrix.  The objective of LSM is to minimize ${\bf{P}}({\bf{x}})$. Since the second derivative of ${\mathbf{P}(\mathbf{x})}$ is a positive definite matrix, ${\mathbf{P}(\mathbf{x})}$ is minimized when the derivative equals zero. Then, the estimation error $\delta {\bf{x}}$ can be defined as
\begin{equation}\label{delta x}
\delta {\bf{x}}  = - {\left( {{{\bf{A}}^{\rm{T}}}{\bf{W_e A}}} \right)^{ - 1}}{{\bf{A}}^{\rm{T}}}{\bf{W_e n}}.
\end{equation}
where $\delta {\bf{x}}$ is solely dependent on $\bf{A}$ and $\bf{n}$. When $\bf{n}$ follows a Gaussian distribution with a mean of $0$, the estimation of LSM is unbiased. 

\subsection{Estimate AoA}
To estimate the AoA of signal from TIS array $r$ to user, we first define the transmit array factor ${{\bf{a}}_t}\left(  \cdot  \right)$ and receive array factor ${{\bf{a}}_r}\left(  \cdot  \right)$ as follow
\begin{equation}
{{\bf{a}}_{t/r}}\left( \psi  \right) = \frac{1}{{\sqrt N }}{\left[ {1,{e^{ - j\frac{{2\pi d}}{\lambda }\sin \psi }}, \ldots ,{e^{ - j\frac{{2\pi d}}{\lambda }\left( {N - 1} \right)\sin \psi }}} \right]^{\rm{T}}},
\end{equation}
where $N$ denotes the number of antenna. $d$ denotes the antenna spacing. $\psi$ denotes the offset angle of the transmit/receive signal. With the aid of ${{\bf{a}}_t}\left(  \cdot  \right)$ and ${{\bf{a}}_r}\left(  \cdot  \right)$, ${{h}_{ir,mk}}$ can be further described as
\begin{equation}
{{\bf{h}}_{ir}} = {{\bf{C}}_{ir}} \odot {{\bf{a}}_r}\left( {{\gamma _{sr}}} \right){\bf{a}}_t^{\rm{H}}\left( {\frac{\pi }{2} - {\gamma _{sr}}} \right),
\end{equation}
where ${{\bf{C}}_{ir}}$ is a constant vector. $\odot$ denotes Hadamard product.
Similarly, ${{h}_{ru,kw}}$ can be further described as
\begin{equation}
{{\bf{h}}_{ru}} = {{\bf{C}}_{ru}} \odot {{\bf{a}}_r}\left( {{\gamma _{ru}}} \right){\bf{a}}_t^{\rm{H}}\left( {\frac{\pi }{2} - {\gamma _{ru}}} \right),
\end{equation}
where ${{\bf{C}}_{ru}}$ is a constant vector. Then, the received signal from all directions ${{\bf{y}}_r}$ can be given by
\begin{equation}
\begin{aligned}
{{\bf{y}}_r} = & {\bf{W}}{{\bf{a}}_r}\left( { {\gamma _{ru}}} \right){\bf{a}}_t^{\rm{H}}\left( {\frac{\pi }{2} - {\gamma _{ru}}} \right){{\bf{\Psi }}_r}{{\bf{a}}_r}\left( { {\gamma _{sr}}} \right){\bf{a}}_t^{\rm{H}}\left( {\frac{\pi }{2} -{\gamma _{sr}}} \right)\\
&\sum\limits_{i = 1}^I {\left( {\sqrt {{L_{ir}}{L_{ru}}{P_{\rm{T}}}} {{\bf{C}}_{ir}} \odot {{\bf{C}}_{ru}} \odot {{\bf{p}}_i}} \right)}{s_{ir}} + {{\bf{n}}_0},
\end{aligned}
\end{equation}
where ${\bf{W}} = \left[ {{{\bf{w}}_1}, \ldots ,{{\bf{w}}_D}} \right]$ denotes the user codebook. $D$ denotes the number of directions. ${{\bf{n}}_0}$ denotes the noise vector from all directions. The concerned ${\hat \gamma }$ is estimated utilizing ML estimator as
\begin{equation}\label{AoA_opt}
{{\hat \gamma }_{ru}} = \mathop {{\mathop{\rm argmin}\nolimits} }\limits_{\theta ,\phi  \in [0,\pi /2]} {\left\| {{{\bf{y}}_r} - {\bf{D}}({\gamma _{ir}},{\gamma _{ru}}){{{\bf{D}}^{\rm{M}}}}({\gamma _{ir}},{\gamma _{ru}}){{\bf{y}}_r}} \right\|^2},
\end{equation}
where ${{{\bf{D}}^{\rm{M}}}}$ denotes the Moore-Penrose inverse of ${\bf{D}}$. ${\bf{D}}(a,b)$ denotes dictionary matrix, which is given by 
\begin{equation}
\begin{aligned}
{\bf{D}}(a,b) = &{\bf{W}}{{\bf{a}}_r}\left( b \right){\bf{a}}_t^{\rm{H}}\left( {\frac{\pi }{2} - b} \right){{\bf{\Psi }}_r}{{\bf{a}}_r}\left( a \right){\bf{a}}_t^{\rm{H}}\left( {\frac{\pi }{2} - a} \right)\\
&\sum \limits_{i = 1}^I {\left( {\sqrt {{L_{ir}}{L_{ru}}{P_{\rm{T}}}} {{\bf{C}}_{ir}} \odot {{\bf{C}}_{ru}} \odot {{\bf{p}}_i}} \right)}.
\end{aligned}
\end{equation}
Through the algorithm proposed in \cite{AoA}, we can solve the optimization problem \eqref{AoA_opt} and obtain the AoA.

\subsection{Positioning user}

Based on the positions of the TIS arrays and the AoA, we can get an expression for a line. Ideally, the intersection of all lines is the user's location. Considering measurement errors, we relax the intersection points to the one with the smallest total distance from the $R$ lines. Due to the noise and inherent error, $R$ rays may not intersect. Therefore, we propose the following optimization objective function:
\begin{equation}\label{optimization objective function1}
\begin{array}{*{20}{c}}
{\mathop {\min }\limits_{{x_u},{y_u},{z_u}} }&{\sum\limits_{r = 1}^R {{d_r}\left( {{x_u},{y_u},{z_u}} \right)} .}
\end{array}
\end{equation}
where ${{d_r}\left( {{x_u},{y_u},{z_u}} \right)}$ denotes the distance between the user and TIS array $r$.

Subsequently, various optimization techniques including the mean value method (MVM), nonlinear unconstrained optimization method (NUOM), LSM and gradient descent method (GDM) are applied to determine the positioning results which minimize the optimization objective function. In MVM, we only consider positioning in two dimensions in MVM algorithm. Ignoring the possibility of parallelism, any two lines will have a point of intersection, resulting in ${\rm{C}}_R^2$ intersection points.

TSIPA offers a significant advantage: the positioning process in the second stage operates independently of satellite involvement. Once the positions of the TIS arrays are established in the first stage, the user can autonomously perform positioning without requiring additional satellite communication. The above three stages constitute the whole process of TSIPA, which is shown in{\textbf{ Algorithm~\ref{TSIPA}}}. $K_{max}$ denotes the max number of iterations.
\begin{algorithm} 
\caption{TSIPA}
\label{TSIPA}
\begin{algorithmic}[1]
\STATE \textbf{Positioning for the first time}
\STATE \textbf{1st stage:} locate the TIS arrays
\STATE \textbf{Input:} ephemeris information
\STATE Determine the positions of TIS arrays.

\STATE \textbf{2nd stage:} estimate AoA
\STATE \textbf{Input:} the positions of the TIS arrays
\STATE Obtain AoA through maximum likelihood estimation

\STATE \textbf{3rd stage:} locate the user 
\STATE \textbf{Input:} the positions of the TIS arrays and AoA
\STATE Resolve problem \eqref{optimization objective function1}
\STATE \textbf{Positioning after the first time}
\STATE \textbf{Input:} the locations of TIS arrays obtained in the first time positioning and estimate AoA only through TIS
\STATE Determine the position of user through LSM.
\end{algorithmic}
\end{algorithm}
In this section, we analyze the error from two perspectives: the position of the TIS arrays and the position of the user. The main objective is to investigate what factors influence the positioning error of user obtained through TSIPA.

To evaluate the accuracy of TIS positioning, the dilution of precision (PDoP) can be determined by computing the parameter covariance matrix. Similarly, to evaluate the distribution of TIS arrays, we define a novel navigation performance evaluation metric called TPDoP, to determine the distance between the estimated user location and the centroid of the TIS arrays. TPDoP is given by:
\begin{equation}\label{PDoP}
{\rm{TPDoP}} = \sum\limits_{r = 1}^R {\sqrt {{{\left( {{x_r} - {x_c}} \right)}^2} + {{\left( {{y_r} - {y_c}} \right)}^2} + {{\left( {{z_r} - {z_c}} \right)}^2}} } ,
\end{equation}
where ${x_c}$, ${y_c}$ and ${z_c}$ are the coordinates of the geometric centers of the shapes enclosed by all TIS arrays. 

Although TPDoP can represent the degree of centroid deviation of TIS arrays from the user, the compactness of TIS arrays is still unknown. Then, we utilize RMSE to characterize the compactness, where the RMSE of TIS arrays positions can be given by
\begin{equation}\label{PDoP}
{\rm{RMSE}} = \sqrt {\frac{1}{R}\sum\limits_{r = 1}^R {{{\left( {{d_{r,\rm{TPDoP}}} - \bar d} \right)}^2}} } 
\end{equation}
where ${d_{r,\rm{TPDoP}}}$ denotes the distance between TIS array $r$ and the user. $\bar d$ denotes the average distance between each TIS array and the user. By utilizing RMSE, we can derive if one or more TIS arrays are too far away from the user. 

Angular ambiguity is a critical factor in the second stage of TSIPA. Our focus is primarily on the beamwidth emitted by the TIS and the resolution of AoA detection. By utilizing the well-established MIMO technology, the half-power beamwidth (HPBW) of a planar array antenna in the elevation plane can be expressed as:
\begin{equation}\label{beamwidth_theta}
{\theta _h} \approx \frac{1}{{\cos {\theta _0}\sqrt {\Delta \theta _x^{ - 2}{{\cos }^2}{\phi _0} + \Delta \theta _y^{ - 2}{{\sin }^2}{\phi _0}} }},
\end{equation}
where ${{\theta _0}}$ denotes the elevation angles of the beam. ${\Delta \theta _x}$ and ${\Delta \theta _y}$  denote the HPBW of the x-axis and y-axis linear array antennas, respectively. 
Since the TIS array has equal length and width, it can be modeled as a uniform planar array antenna, where the HPBW for elevation can be expressed as follows:
\begin{equation}\label{bw_UPA_theta}
{\theta _h} = \frac{{\Delta {\theta _x}}}{{\cos {\theta _0}}} = \frac{{\Delta {\theta _y}}}{{\cos {\theta _0}}} = \frac{{102\lambda }}{{\sqrt K \zeta \cos {\theta _0}}},
\end{equation}
where $\zeta$ denotes the element spacing, and $102$ is an empirical coefficient that accounts for the effect of the antenna array sidelobe. \eqref{bw_UPA_theta} is expressed in degrees. 

The angle ambiguity, which is numerically equivalent to the HPBW value, is influenced by the user's position, thereby affecting the positioning accuracy. It is evident that when the angular ambiguity increases, the user range will increase thereby, resulting in worse positioning accuracy. Based on the insights from \eqref{bw_UPA_theta}, the user's orientation relative to the TIS array affects the positioning accuracy.

\section{Numerical Results}

We conducted an extensive numerical analysis to evaluate the performance characteristics of TSIPA in a TIS-aided satellite indoor positioning system. The accuracy of the theoretical model is validated through Monte Carlo simulations.

\begin{figure}[ht]
\centering
\includegraphics[width =0.95\columnwidth]{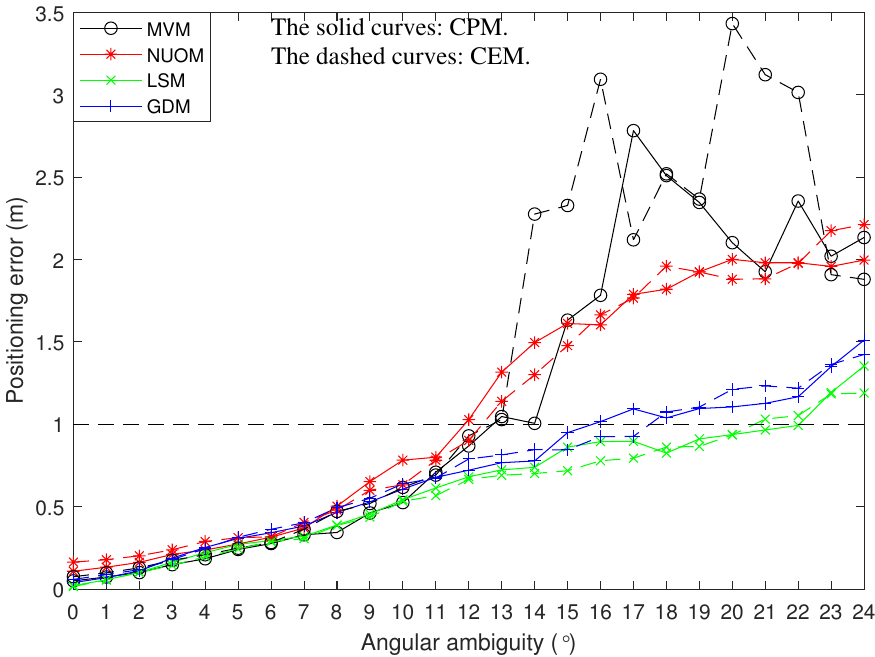}
\caption{The positioning error versus angular ambiguity in CEM and CPM.}
\label{figure_angular_ambiguity}
\end{figure}

In Fig. \ref{figure_angular_ambiguity}, we evaluate the positioning errors in both CEM and CPM. Positioning errors are calculated across $25$ distinct levels of angular ambiguity while keeping the satellites, TIS arrays, and user positions fixed. In Fig. \ref{figure_angular_ambiguity}, the positioning error of TIS arrays is approximately $1$ m for CEM and $0.7$ m for CPM. For both CEM and CPM, the positioning errors increase as the angular ambiguity rises. CPM yields smaller positioning errors compared to CEM. Both CEM and CPM exhibit similar trends when angular ambiguity is under $7^\circ$. The positioning performance of the both indicates that the positioning accuracy of the TIS arrays has a consistent impact on the positioning errors introduced by angular ambiguity. In addition, the LSM demonstrates the lowest positioning error. More specifically, the curves of MVM show poorer stability compared to LSM, with a rapid increase in positioning error and noticeable jitter when angular ambiguity exceeds approximately \( 11^\circ \). The curves of NUOM show a faster rise compared to LSM in positioning error around \( 7^\circ \). The trends for the curves of LSM and GDM are similar, but the positioning error of GDM is about $10$ cm higher than that of LSM. At an angular ambiguity of approximately \( 22^\circ \), corresponding to a TIS elements dimension of \( 9 \times 9 \) units, the errors for LSM reach \( 1 \, \text{m} \). The variation in the number of TIS elements shows the superiority of LSM applied to TSIPA.

\begin{figure}[ht]
\centering
\includegraphics[width =0.95\columnwidth]{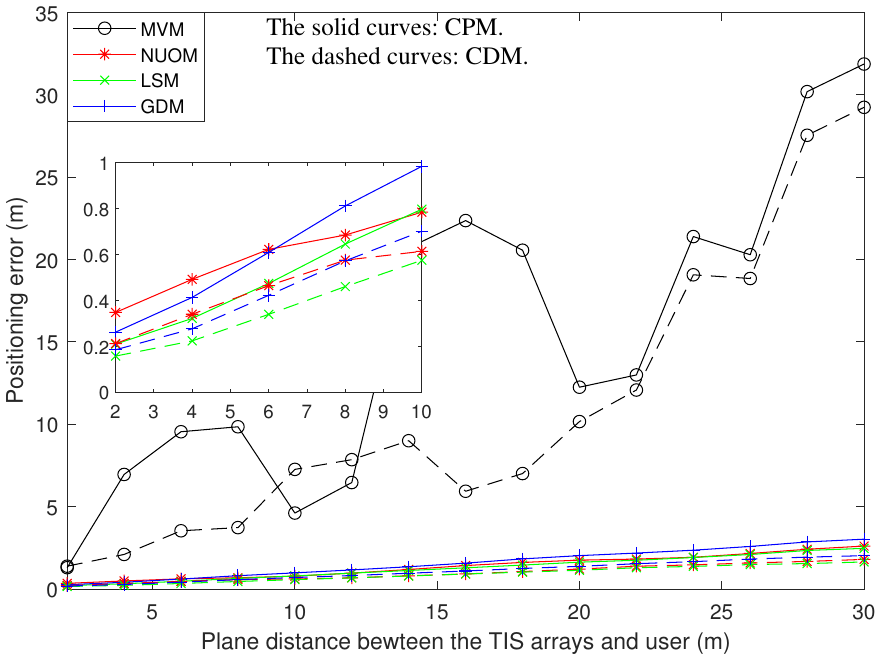}
\caption{The positioning error versus the distance between the TIS arrays and user in CEM and CPM.}
\label{figure_B}
\end{figure}

As shown in Fig. \ref{figure_B}, we evaluate the positioning error relative to the plane distance between the TIS arrays and the user in both CEM and CPM. To control variables, we fix the user's position with TIS arrays distributed at  \(0^\circ\), \(30^\circ\), \(60^\circ\), and \(90^\circ\) in the horizontal plane. The $z$-axis coordinates remain identical to the user's position. The angular ambiguity of TIS arrays is set to \(7.5^\circ\). As the distance between the TIS arrays and the user increases, the positioning error also increases accordingly. Specifically, the curves of MVM exhibit the largest and most unstable positioning errors in both CEM and CPM. The curves of GDM shows positioning errors approximately $30$ cm larger than those of NUOM and LSM. The positioning errors of both NUOM and LSM are nearly identical and the error for LSM remains slightly smaller than that of NUOM. Based on the quantitative results, the positioning error of LSM is the minimum.

\begin{figure}[ht]
\centering
\includegraphics[width = 0.95\columnwidth]{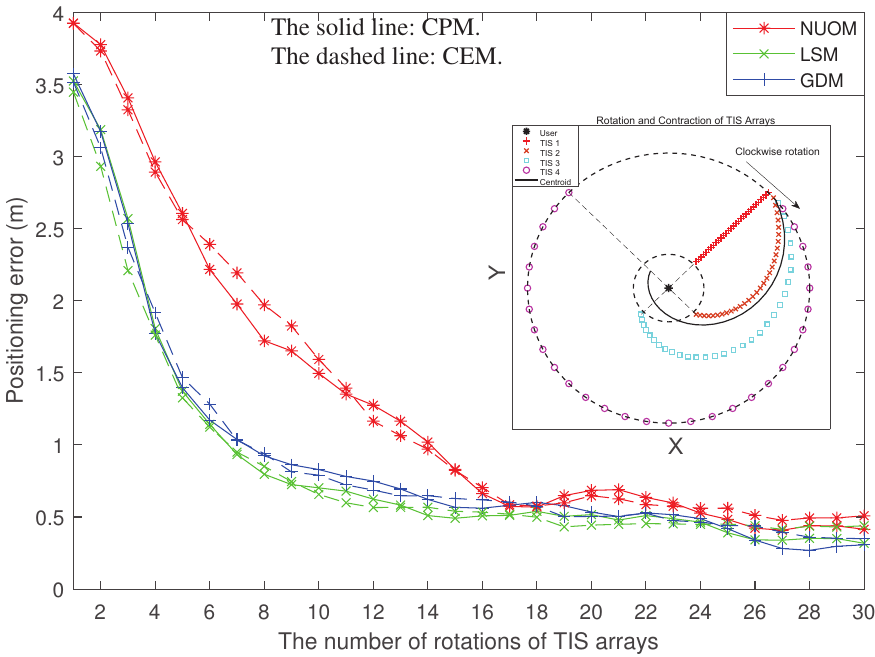}
\caption{The relationship between performance metrics and the number of rotations in CEM and CPM.}
\label{figure_C_ana}
\end{figure}

In Fig. \ref{figure_C_ana}, four TIS arrays are arranged in the x-y plane, with their initial positions set at a same location. To adjust the positions of the TIS arrays and realize the simulation of TIS around the user's rotation process, we assume that: the first TIS array remains stationary, while the second, third, and fourth TIS arrays rotates clockwise in steps of $3^\circ$, $6^\circ$, and $9^\circ$ per turn. Finally, the four TIS arrays are located in four perpendicular directions relative to the user. As the rotations progress, their spatial distribution gradually shifts from a compact to a more dispersed configuration. Meanwhile, the TIS array $1$, $2$, and $3$ also approaches the user as it rotates, while TIS array $4$ remains a constant distance to user at $4$ meters. The black curve in Fig. \ref{figure_C_ana} represents the movement trajectories of the centroid of the TIS arrays.

The relationship between the TPDoP, RMSE, and the distribution of TIS arrays in CEM and CPM is shown in Fig. \ref{figure_C_ana}. As the TIS arrays gradually rotate to mutually perpendicular directions, the curves of TPDoP exhibits a decreasing trend. For RMSE, TIS array $4$ is farther away from the circle formed by the other three TIS arrays, resulting in RMSE increasing with motion, which accords with the expectation of theoretical deduction. The curves of CPM and CEM are closely intertwined, exhibiting only tiny differences, which demonstrates that the positioning errors of TIS arrays have a negligible impact on the TPDoP.

\section{Conclusion}

In this paper, a novel satellite indoor positioning method of TIS-aided satellite indoor navigation system was investigated. Initially, we outlined the existing challenges faced by current indoor positioning technologies and expounded the ability of TIS to introduced ELoS links. We proposed the TSIPA to solve the indoor positioning problem in two stages, and provide the user with subsequent indoor positioning services without communicating with satellites again. We evaluated the positioning performance of CPM and CEM in the same scenario leading to the conclusion that CPM is slightly superior. Then, we analyzed the variation of user positioning error with TPDoP and angle ambiguity. LSM is considered to be the optimal method. The results obtained in this paper confirm the effectiveness of employing TISs for providing satellite indoor positioning services, which motivates related future research on TISs, such as low earth orbit (LEO) satellite navigation and INAC. 

\bibliographystyle{IEEEtran}
\bibliography{IEEEabrv,bib2014}

\begin{thebibliography}{10}
\providecommand{\url}[1]{#1}
\csname url@samestyle\endcsname
\providecommand{\newblock}{\relax}
\providecommand{\bibinfo}[2]{#2}
\providecommand{\BIBentrySTDinterwordspacing}{\spaceskip=0pt\relax}
\providecommand{\BIBentryALTinterwordstretchfactor}{4}
\providecommand{\BIBentryALTinterwordspacing}{\spaceskip=\fontdimen2\font plus
\BIBentryALTinterwordstretchfactor\fontdimen3\font minus
  \fontdimen4\font\relax}
\providecommand{\BIBforeignlanguage}[2]{{%
\expandafter\ifx\csname l@#1\endcsname\relax
\typeout{** WARNING: IEEEtran.bst: No hyphenation pattern has been}%
\typeout{** loaded for the language `#1'. Using the pattern for}%
\typeout{** the default language instead.}%
\else
\language=\csname l@#1\endcsname
\fi
#2}}
\providecommand{\BIBdecl}{\relax}
\BIBdecl

\bibitem{INT_GNSS1}
A.~G. Dempster and E.~Cetin, ``{Interference Localization for Satellite
  Navigation Systems},'' \emph{Proc. IEEE}, vol. 104, no.~6, pp. 1318--1326,
  Jun. 2016.

\bibitem{INT_GNSS3}
A.~Yassin, Y.~Nasser, M.~Awad, A.~Al-Dubai, R.~Liu, C.~Yuen, R.~Raulefs, and
  E.~Aboutanios, ``{Recent Advances in Indoor Localization: A Survey on
  Theoretical Approaches and Applications},'' \emph{IEEE Commun. Surv. Tutor.},
  vol.~19, no.~2, pp. 1327--1346, Second quarter 2017.

\bibitem{INT_GNSS_indoor}
T.~Janssen, A.~Koppert, R.~Berkvens, and M.~Weyn, ``{A Survey on IoT
  Positioning Leveraging LPWAN, GNSS, and LEO-PNT},'' \emph{IEEE Internet
  Things J.}, vol.~10, no.~13, pp. 11\,135--11\,159, Jul. 2023.

\bibitem{INT_indoor2}
P.~Puricer and P.~Kovar, ``{Technical Limitations of GNSS Receivers in Indoor
  Positioning},'' in \emph{2007 17th International Conference
  Radioelektronika}, Brno, Czech Republic, Apr. 2007, pp. 1--5.

\bibitem{INT_LoS}
J.~Blanch, T.~Walter, and P.~Enge, ``{Position error bound calculation for GNSS
  using measurement residuals},'' \emph{IEEE Trans. Aerosp. Electron. Syst.},
  vol.~44, no.~3, pp. 977--984, Jul. 2008.

\bibitem{INT_TIS1}
Y.~Liu, X.~Liu, X.~Mu, T.~Hou, J.~Xu, M.~Di~Renzo, and N.~Al-Dhahir,
  ``{Reconfigurable Intelligent Surfaces: Principles and Opportunities},''
  \emph{IEEE Commun. Surv. Tutor.}, vol.~23, no.~3, pp. 1546--1577, third
  quarter 2021.

\bibitem{INT_STARS1}
X.~Mu, Y.~Liu, L.~Guo, J.~Lin, and R.~Schober, ``{Simultaneously Transmitting
  and Reflecting (STAR) RIS Aided Wireless Communications},'' \emph{IEEE Trans.
  Wireless Commun.}, vol.~21, no.~5, pp. 3083--3098, May 2022.

\bibitem{INT_ELoS}
Q.~Zhao, W.~Gong, T.~Hou, X.~Sun, and E.~Bodanese, ``{Global Navigation
  Satellite System (GNSS): A Reconfigurable Intelligent Surface (RIS)-aided
  Approach},'' in \emph{GLOBECOM 2022 - 2022 IEEE Global Communications
  Conference}, Rio de Janeiro, Brazil, Dec. 2022, pp. 3162--3167.

\bibitem{INT_INAC1}
T.~Hou and A.~Li, ``{Performance Analysis of NOMA-RIS Aided Integrated
  Navigation and Communication (INAC) Networks},'' \emph{IEEE Trans. Veh.
  Technol.}, vol.~72, no.~10, pp. 13\,255--13\,268, Oct. 2023.

\bibitem{INT_INAC2}
Q.~Zhao, W.~Gong, T.~Hou, X.~Sun, A.~Li, and E.~Bodanese,
  ``{Integrated-Navigation-and-Communication (INAC): A Reconfigurable
  Intelligent Surface (RIS)-aided Approach},'' in \emph{2023 IEEE 97th
  Vehicular Technology Conference (VTC2023-Spring)}, Florence, Italy, Jun.
  2023, pp. 1--6.

\bibitem{shadowed_rice}
J.~R. Gangane, M.~C. Aguayo-Torres, and J.~J. Sanchez-Sanchez, ``{Performance
  of SIMO MRC SC-FDMA over shadowed Rice Land Mobile Satellite channel},'' in
  \emph{Wireless VITAE 2013}, Atlantic City, NJ, USA, Jun. 2013, pp. 1--5.

\bibitem{LAMBDA}
N.~Nadarajah, P.~J.~G. Teunissen, and N.~Raziq, ``{Instantaneous {GPS-Galileo}
  Attitude Determination: Single-Frequency Performance in Satellite-Deprived
  Environments},'' \emph{IEEE Trans. Veh. Technol.}, vol.~62, no.~7, pp.
  2963--2976, Sep. 2013.

\bibitem{AoA}
N.~Varshney and S.~De, ``{AoA-Based Low Complexity Beamforming for Aerial RIS
  Assisted Communications at mmWaves},'' \emph{IEEE Commun. Lett.}, vol.~27,
  no.~6, pp. 1545--1549, Jun. 2023.

\end{thebibliography}

\end{document}